\documentclass[]{emulateapj}

\usepackage{latexsym}

\shorttitle{Discovery of Extreme Carbon Stars in the LMC}
\shortauthors{Gruendl et al.}

\begin{document}

\title{Discovery of Extreme Carbon Stars in the Large Magellanic Cloud}

\author{R.~A.\ Gruendl\altaffilmark{1,2},
  Y.-H.\ Chu\altaffilmark{1,2}, 
  J.~P.\ Seale\altaffilmark{1}, 
  M.\ Matsuura\altaffilmark{3,4},
  A.~K.\ Speck\altaffilmark{5}, 
  G.~C.\ Sloan\altaffilmark{6}, 
  L.~W.\ Looney\altaffilmark{1}} 
\altaffiltext{1}{Astronomy Department, University of Illinois, 
        1002 W. Green Street, Urbana, IL 61801;
        gruendl@astro.uiuc.edu, chu@astro.uiuc.edu, seale@astro.uiuc.edu, 
        lwl@astro.uiuc.edu}
\altaffiltext{2}{Visiting astronomer, Cerro Tololo Inter-American Observatory}
\altaffiltext{3}{National Astronomical Observatory of Japan, Osawa 2-21-1, Mitaka, 
Tokyo 181-8588, Japan; m.matsuura@nao.ac.jp}
\altaffiltext{4}{Department of Physics and Astronomy, University College London,
Gower Street, London WC1E 6BT, UK}
\altaffiltext{5}{Department of Physics \& Astronomy, University of Missouri, Columbia, 
MO 65211; speckan@missouri.edu}
\altaffiltext{6}{Astronomy Department, Cornell University, 610 Space Sciences Building, 
Ithaca, NY 14853-6801, USA}

\begin{abstract}
Using {\it Spitzer} IRAC and MIPS observations of the Large Magellanic Cloud, we have 
identified 13 objects that have extremely red mid-IR colors.  Follow-up {\it Spitzer} 
IRS observations of seven of these sources reveal varying amounts of SiC and 
C$_2$H$_2$ absorption as well as the presence of a broad MgS feature in at least two 
cases, indicating that these are extreme carbon stars.
Preliminary estimates find these objects have luminosities of 4--11$\times$10$^3$~$L_\odot$
and preliminary model fitting gives mass-loss rates between 4$\times$10$^{-5}$ and 
2$\times$10$^{-4}$~$M_\odot$~yr$^{-1}$, higher than any known carbon-rich AGB star
in the LMC.  These spectral and physical properties require careful reconsideration of
dust condensation and mass-loss processes for carbon stars in low metallicity environments.
\end{abstract}


\section{Introduction}

We have undertaken a study of star formation in the Large Magellanic Cloud 
\citep[LMC;][]{GC08}, using archival {\it Spitzer Space Telescope}
observations, such as those of Surveying the Agents of a Galaxy's 
Evolution \citep[SAGE;][]{Metal06}.
In the process of identifying young stellar objects (YSOs) we noticed 13 
bright mid-IR sources (Table 1) that all had similar photometric properties 
with spectral energy distributions (SEDs) markedly different from 
those of evolved stars or YSOs: (1) they have extremely red mid-IR colors,
[4.5]-[8.0]$>$4.0; (2) their SEDs, peaking between 8 and 24~$\mu$m, can be 
moderately well fit by blackbodies with effective temperatures of 
$\sim$230--320~K; (3) they all fall in a narrow range of brightness, 
7.0$>$[8.0]$>$8.5; and (4) none of the sources have counterparts in the Two Micron 
All Sky Survey Point Source Catalog \citep[2MASS PSC;][]{Setal06} or in the Digitized 
Sky Survey.  We dubbed these sources Extremely Red Objects (EROs).  

The EROs are not likely asteroids because of their high brightnesses and 
lack of proper motions in SAGE observations from two epochs separated by $\sim$3 
months.  They cannot be background galaxies or Galactic sources, as the 
{\it Spitzer} Wide-area IR Extragalactic Survey \citep[SWIRE;][]{Letal03} and 
the Galactic Legacy IR Mid-Plane Survey Extraordinaire \citep[GLIMPSE;][]{Betal03}
do not have counterparts with similar mid-IR colors and brightnesses.  These EROs are most 
likely associated with the LMC,
and their high luminosities imply that their Galactic counterparts would have 
saturated in the SWIRE and GLIMPSE Surveys.

Queries of the SIMBAD database found that some EROs had been detected
previously in either {\it MSX} or {\it IRAS} observations \citep[e.g.,][]{S89}.
Five have been suggested as possible obscured AGB stars by \citet{Letal97};
however, many of our EROs have also been suggested to be YSO candidates \citep{Wetal08}.
As these EROs have never been confirmed or rejected as YSOs spectroscopically,
we included seven among our follow-up observations of massive YSOs in the LMC
using the {\it Spitzer} InfraRed Spectrograph \citep[IRS;][]{Hetal04}.
These IRS observations show unambiguously that the EROs are carbon
stars; furthermore, the spectra reveal silicon carbide (SiC) absorption features.
While SiC emission features are fairly common in both Galactic and LMC carbon
stars, this is the first clear detection of SiC {\it absorption} for LMC carbon stars.
Preliminary analysis of the IRS spectra suggests that these are extraordinary carbon 
stars with very high mass-loss rates.  This paper reports their discovery.
In \S{2} we describe their basic photometric properties 
and their possible optical and near-IR counterparts.  In \S{3} we introduce 
our IRS observations and in \S{4} we discuss the results.

\section{Photometric Observations}

\subsection{Mid-Infrared Observations}

\begin{deluxetable*}{lcccccclcc}[]
\tablewidth{0pt}
\tabletypesize{\scriptsize}
\tablecaption{Mid-Infrared Photometric Properties and Derived Parameters\label{tab_ERO}}
\tablehead{
\colhead{} & 
\colhead{} & \colhead{} & \colhead{} & 
\colhead{} & \colhead{} & \colhead{} & 
\colhead{Cross} & \colhead{} & \colhead{Mass-Loss} \\
\colhead{Source ID} & 
\colhead{m$_{3.6}$} & \colhead{m$_{4.5}$} & \colhead{m$_{5.8}$} & 
\colhead{m$_{8.0}$} & \colhead{m$_{24.0}$} & \colhead{m$_{70.0}$} &
\colhead{Identification} & \colhead{L$_{\rm bol}$} & \colhead{Rate} \\
\colhead{ } & 
\colhead{[mag]} & \colhead{[mag]} & \colhead{[mag]} & 
\colhead{[mag]} & \colhead{[mag]} & \colhead{[mag]} &
\colhead{} & \colhead{[$L_\odot$]} & \colhead{[$M_\odot$~yr$^{-1}$]}  
}
\startdata
 050231.49$-$680535.8 & 15.92$\pm$.08  & 12.75$\pm$.05  & \phn9.91$\pm$.05  &  7.26$\pm$.05  &  3.06$\pm$.11 &  2.09$\pm$.23 &                    & \phn7800 & 1.1$\times$10$^{-4}$ \\
 050343.02$-$664456.7 & 14.92$\pm$.06  & 12.93$\pm$.06  &    10.68$\pm$.06  &  8.25$\pm$.06  &  2.96$\pm$.11 &  1.71$\pm$.24 & IRAS\,05036$-$6649 & \phn7000 & 1.5$\times$10$^{-4}$ \\
 050405.60$-$682340.3 & 16.47$\pm$.10  & 13.12$\pm$.06  &    10.27$\pm$.05  &  7.65$\pm$.05  &  3.63$\pm$.11 &   $>$3.32\tablenotemark{a}     & IRAS\,05042$-$6827 & \phn4950 & 6.3$\times$10$^{-5}$ \\
 051301.75$-$693351.0 & 15.87$\pm$.08  & 13.70$\pm$.06  &    10.69$\pm$.06  &  7.78$\pm$.06  &  3.35$\pm$.11 &   $>$1.59\tablenotemark{a}     & IRAS\,05133$-$6937 & \phn5800 & 8.9$\times$10$^{-5}$ \\
 051811.70$-$703027.0 & 14.97$\pm$.07  & 12.26$\pm$.06  & \phn9.89$\pm$.05  &  7.61$\pm$.05  &  2.82$\pm$.11 &  1.59$\pm$.22 & IRAS\,05187$-$7033 & \phn8850 & 1.3$\times$10$^{-4}$ \\
 051848.36$-$693334.7 & 14.41$\pm$.07  & 12.34$\pm$.06  & \phn9.72$\pm$.05  &  7.11$\pm$.05  &  3.32$\pm$.11 &  1.24$\pm$.24 & IRAS\,05191$-$6936 & \phn7750 & 8.3$\times$10$^{-5}$ \\
 052540.63$-$700827.2 & 16.02$\pm$.13  & 13.20$\pm$.06  &    10.45$\pm$.06  &  7.80$\pm$.05  &  3.82$\pm$.11 &  2.58$\pm$.26 & IRAS\,05260$-$7010 & \phn4300 & 5.9$\times$10$^{-5}$ \\
 052937.89$-$724952.9 & 13.61$\pm$.06  & 11.14$\pm$.05  & \phn9.02$\pm$.05  &  7.09$\pm$.05  &  3.81$\pm$.11 &  2.92$\pm$.26 & IRAS\,05305$-$7251 & \phn5550 & 4.2$\times$10$^{-5}$ \\
 053044.10$-$714300.5 & 13.47$\pm$.06  & 12.13$\pm$.05  &    10.21$\pm$.05  &  7.76$\pm$.06  &  2.75$\pm$.11 &  1.13$\pm$.22 & IRAS\,05315$-$7145 & \phn9500 & 1.7$\times$10$^{-4}$ \\
 054134.73$-$694209.3 & 14.84$\pm$.07  & 11.91$\pm$.05  & \phn9.40$\pm$.05  &  7.11$\pm$.05  &  3.50$\pm$.11 &   $>$2.35\tablenotemark{a}     & IRAS\,05420$-$6943 & \phn6400 & 6.6$\times$10$^{-5}$ \\
 054859.98$-$703322.5 & 15.33$\pm$.08  & 13.70$\pm$.06  &    11.32$\pm$.06  &  8.41$\pm$.06  &  2.41$\pm$.11 &  0.72$\pm$.22 & IRAS\,05495$-$7034 &    11100 & 2.3$\times$10$^{-4}$ \\
 055026.08$-$695603.1 & 14.26$\pm$.05  & 11.81$\pm$.05  & \phn9.25$\pm$.05  &  6.94$\pm$.05  &  3.10$\pm$.11 &  2.36$\pm$.23 & IRAS\,05509$-$6956 & \phn8250 & 9.0$\times$10$^{-5}$ \\
 055133.60$-$711933.9 & 14.06$\pm$.06  & 11.28$\pm$.05  & \phn9.03$\pm$.05  &  7.01$\pm$.05  &  3.87$\pm$.11 &  3.38$\pm$.26 & IRAS\,05522$-$7120 & \phn5200 & 4.4$\times$10$^{-5}$   
\enddata
\tablenotetext{a}{For non-detections a 3-$\sigma$ upper-limit is given.}
\end{deluxetable*}

\begin{figure}[]
\plotone{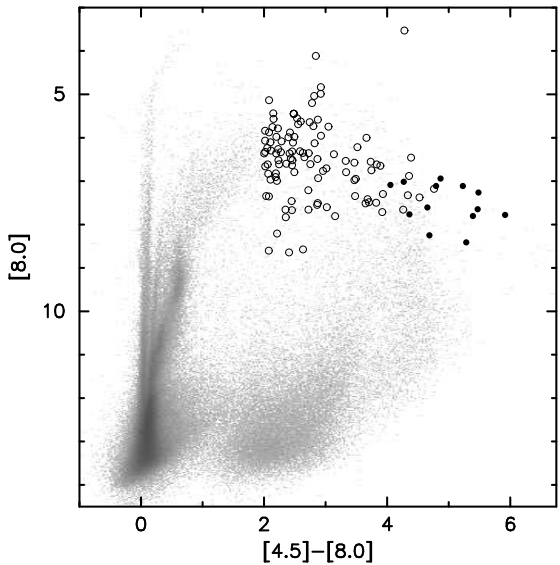}
\caption{Color-magnitude diagram showing IRAC [8.0] vs. [4.5]$-$[8.0].  The gray scale
Hess diagram shows all data from the LMC.  The location of the EROs are plotted with
filled circles while other sources classified as evolved stars by \citet{GC08}
have been plotted with open circles.}
\label{ERO_CMD}
\end{figure}

{\it Spitzer} observations of the EROs, from SAGE \citep{Metal06}, 
were made with the InfraRed Array Camera \citep[IRAC;][]{Fetal04} and the 
Multiband Imaging Photometer for {\it Spitzer} \citep[MIPS;][]{Retal04}. 
Aperture photometry of the EROs was obtained for the IRAC 3.6, 4.5, 5.8, 
and 8.0~$\mu$m bands and the MIPS 24 and 70~$\mu$m bands using the 
task {\it phot} in IRAF.  A detailed description of the data reduction,
photometric extraction, and uncertainties can be found in \citet{GC08}.
When multiple observations were available, the photometric measurements
were averaged.  Table~\ref{tab_ERO} presents
the average photometric results: column 1 lists source names that
provide epoch 2000 coordinates, columns 2--7 give
flux densities and uncertainties at 3.6, 4.5, 5.8, 
8.0, 24, and 70~$\mu$m, and column 8 gives cross-identifications to 
previously known sources.  
Figure~\ref{ERO_CMD} plots the locations of the EROs in a mid-IR [8.0]
vs. [4.5]$-$[8.0] color-magnitude diagram along with other red sources 
classified as evolved stars by \citet{GC08}.  The EROs 
are at the extreme red end of the tail formed by the evolved stars.

\subsection{Complementary Near-Infrared and Optical Photometry}\label{sec_obsoptnir}

None of the EROs have a counterpart in the 2MASS PSC.  Thus, we obtained deeper
near-IR observations at $J$ and $K_s$-bands for six of the EROs using the IR 
Side Port Imager \citep[ISPI;][]{vdBetal04} on the Blanco 4m telescope at the 
Cerro Tololo Inter-American Observatory in 2007 February.
The resulting images have a typical effective exposure time of $\sim$300~s
and 600~s in the $J$ and $K_s$-bands, respectively, and are flux-calibrated 
using stars in the 2MASS PSC.  The angular resolution of the 
images is typically $\lesssim$~1\farcs0 and the astrometric accuracy is better 
than 0\farcs2.  
For a more detailed description of the near-IR data reduction, see \citet{GC08}.

Table~\ref{tab_NIRsrc} presents the near-IR photometry and 3-$\sigma$ upper limits 
for the six EROs observed.  Only two EROs, 051301.75$-$693351.0 and 
053044.10$-$714300.5, have near-IR counterparts projected within 0\farcs2
as indicated in the last column of the table.
To supplement our ISPI observations, we have used the Magellanic Clouds
Point Source Catalog from the InfraRed Survey Facility \citep[IRSF;][]{Ketal07}.
This survey has an astrometric accuracy of $\sim$0\farcs1 and 90\% completeness 
limits of 18.5 and 17.4 for m$_J$ and m$_K$, respectively. 
IRSF sources with spatial coincidence of 1\arcsec\ or better are also included 
in Table~\ref{tab_NIRsrc}.  Only two IRSF sources are within 0\farcs2 or less 
from their respective EROs and are the same counterparts identified by our ISPI 
observations.  The other three IRSF sources are included for completeness but 
are unlikely to be near-IR counterparts to the EROs.

\begin{deluxetable}{lccr}
\tablecolumns{4}
\tablewidth{0pt}
\tabletypesize{\scriptsize}
\tablecaption{Possible Near-Infrared Counterparts\label{tab_NIRsrc}}
\tablehead{
\colhead{Source ID} & \colhead{m$_{\rm J}$} & \colhead{m$_{\rm K}$} & \colhead{Offset}    \\
\colhead{ }         & \colhead{[mag]}       & \colhead{[mag]}       & \colhead{[\arcsec]} \\
\hline 
\multicolumn{4}{c}{ISPI Observations}
}
\startdata
 050231.49$-$680535.8 &    $>$19.67    &  $>$18.63       &  \\
 050405.60$-$682340.3 &    $>$19.84    &  $>$18.60       &  \\
 051301.75$-$693351.0 & 16.91$\pm$0.05\phm{\,}           & 16.65$\pm$0.06\phm{\,}           & $<$0.2 \\
 052937.89$-$724952.9 &    $>$19.52    &  $>$18.74       &  \\
 053044.10$-$714300.5 & 17.79$\pm$0.07\phm{\,}           & 16.77$\pm$0.07\phm{\,}           & $<$0.2 \\
 055026.08$-$695603.1 &    $>$19.82    &  $>$18.74       &  \\
\cutinhead{Possible Counterparts in IRSF Catalog}
 051301.75$-$693351.0 & 16.98$\pm$0.02\phm{\,}           & 16.27$\pm$0.06\phm{\,}           & 0.1 \\
 051848.36$-$693334.7 & 17.27$\pm$0.03\tablenotemark{a}  & 16.66$\pm$0.12\tablenotemark{a}  & 1.0 \\
 053044.10$-$714300.5 & 18.54$\pm$0.08\phm{\,}           & 17.23$\pm$0.22\phm{\,}           & 0.2 \\
 052540.63$-$700827.2 & 17.77$\pm$0.04\tablenotemark{a}  & 17.08$\pm$0.17\tablenotemark{a}  & 0.7 \\
 054134.73$-$694209.3 & 19.61$\pm$0.21\tablenotemark{a}  & 16.70$\pm$0.09\tablenotemark{a}  & 0.6 
\enddata
\tablenotetext{a}{Doubtful counterpart}
\end{deluxetable}

\begin{figure}[]
\epsscale{1.0}
\plotone{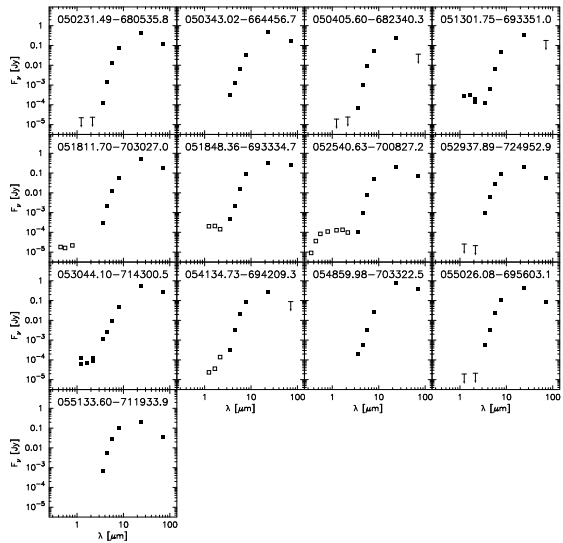}
\caption{The SEDs of the EROs using optical photometry from the 
MCPS, near-IR photometry from our ISPI observations or the IRSF survey, and mid-IR 
photometry from IRAC and MIPS measurements.  Measurements with good astrometric 
coincidence are plotted with filled symbols, while those with poorer coincidence are 
indicated by open symbols.  Upper limits for non-detections are shown
as arrows with a crossbar at the 3-$\sigma$ significance limit.}
\label{fig_SEDs}
\end{figure}

To search for optical counterparts, we used the optical photometry from the Magellanic 
Clouds Photometric Survey \citep[MCPS;][]{Zetal04} which has an astrometric
accuracy of $\sim$0\farcs5 and is generally complete to m$_V=$20.  We find that 
three EROs have faint (m$_V>$19) optical sources within 1$''$, but are likely 
unrelated as none have a spatial coincidence better than 0\farcs7.
In Figure~\ref{fig_SEDs} we combine all photometric data for each source to get a 
broad SED extending from the optical to 70~$\mu$m.

\section{IRS Observations}

Seven EROs were included in our {\it Spitzer} IRS survey of YSOs (PID=40650), 
which used the Short-Low (SL) modules to obtain spectra over $\sim$5.2--14.5~$\mu$m 
with spectral resolution $R$ ($\equiv \delta\lambda/\lambda$) of 64--128, and 
the Short-High (SH) and Long-High (LH) modules to obtain spectra over 
$\sim$9.9--37.2~$\mu$m with spectral resolution of $\sim$600.  Background 
observations with the same integration times were made in nearby regions with 
low surface brightnesses in the {\it IRAS} 12 and 25~$\mu$m maps.

To extract spectra, we used the basic calibrated data (BCD) from 
the {\it Spitzer} Science Center's pipeline.  Rogue pixels and flagged 
data were cleaned using the IRSCLEAN package (ver 1.9).  
Multiple exposures for each slit position were median averaged.  
The background was then removed in each module by subtracting
the cleaned BCD image data at the background position.
For the SL modules, an additional local background was subtracted by 
differencing the two nod positions.  
Spectra were extracted using the SMART software \citep{Higdon04} with a 
full aperture extraction for the SH and LH modules and the tapered column
point source method for the SL modules.
The SH and LH spectra were automatically defringed using the IRSFRINGE package
(ver 1.1).  To combine all spectra, we used the SL extracted spectra to set the 
flux level (as their background subtraction was more robust) and then applied
a multiplicative scale factor for the SH and LH spectra based on the 
continuum where the spectra overlapped.  The resulting spectra span the wavelength 
range from 5.2 to 37.2~$\mu$m.  A more detailed description 
is presented in \citet{Setal09} which reports the results of our IRS survey 
of LMC YSO candidates.

\section{Results}

\begin{figure}[]
\epsscale{1.2}
\plotone{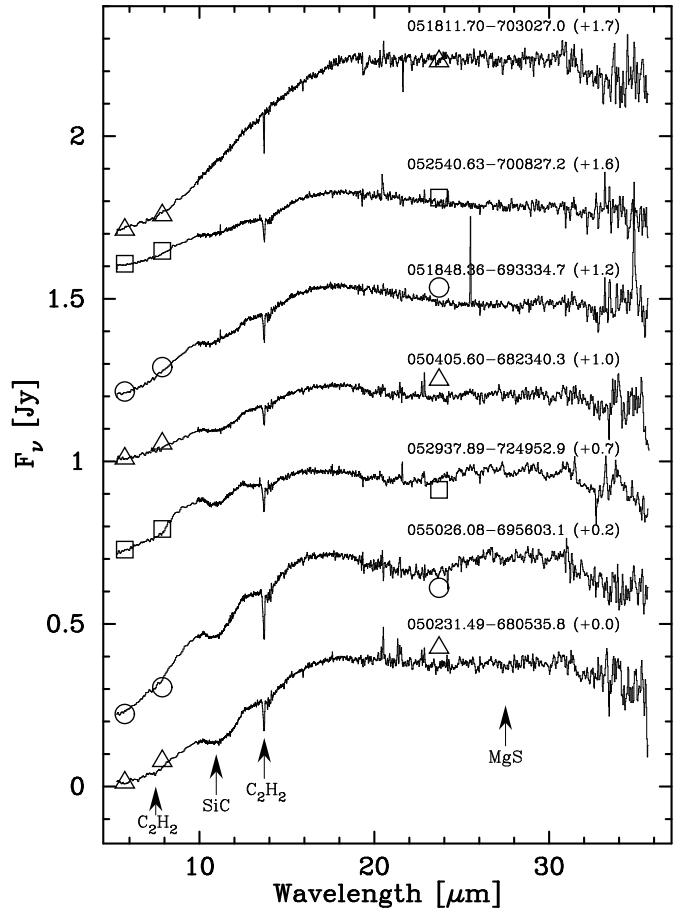}
\caption{IRS spectra for seven EROs.  The spectra are roughly ordered from top to bottom
based on the increasing strength of the SiC feature at 11.3~$\mu$m.  Each spectrum has been 
offset by a constant value in Jy which is given after the source name in the figure.  
Our mid-IR photometry for each source are plotted with each spectrum for comparison
using alternating symbols.}
\label{fig_IRSdata}
\end{figure}

Figure~\ref{fig_IRSdata} presents the IRS spectra of seven EROs along with 
the IRAC and MIPS mid-IR photometry.  Clearly, these spectra
are dominated by dust continuum emission.  Interestingly, the 11.3~$\mu$m SiC
feature is detected in {\it absorption} in nearly all spectra; in 
Figure~\ref{fig_IRSdata} the spectra have been ordered from top to bottom based on 
the strength of the SiC absorption feature.
In addition, the C$_2$H$_2$ (acetylene) 13.7~$\mu$m absorption feature appears
in every spectrum, and the C$_2$H$_2$ 7.5~$\mu$m absorption feature appears
in some spectra.  In contrast, the MgS emission feature at $\sim$26~$\mu$m is 
clearly detected in the spectrum of 055026.08$-$695603.1 and possibly in that of 
052937.89$-$724952.9.  Finally, we note that the only source without 
SiC absorption, 051811.70$-$703027.0, has a nearly featureless spectrum similar 
to AFGL\,618 (the Westbrook Nebula) a carbon-rich protoplanetary nebula 
\citep{Ketal02,Cetal01a,Cetal01b}.

\subsection{Mass-Loss Rates}

The extreme red colors of these EROs, [8.0]$-$[24]= 3--5, suggest 
that they have high opacities and hence high mass-loss rates, $\dot{M}$.
Simple fits to the mid-IR SEDs of the EROs give bolometric luminosities in the range 
of 4--11$\times$10$^3$~$L_\odot$ \citep[for a distance of 50~kpc;][]{F99}.
Preliminary analysis using the radiative transfer model DUSTY \citep{IE95} shows
that the highest optical depths ($\tau$) are found for 054859.98$-$703322.5, with $\tau$'s 
of 270, 8.1, and 10.0 at 1, 10 and 11.3 \,$\mu$m, respectively.  
Assuming a gas-to-dust ratio of 200, the estimated $\dot{M}$ are in the range of 
0.4--$2.3\times10^{-4}$~$M_\odot$~yr$^{-1}$ (see Table~\ref{tab_ERO}).
We have adjusted values of $\dot{M}$ from other works to reflect a gas-to-dust 
ratio of 200.

Our derived values for $\dot{M}$ are higher than previously known for
carbon-rich AGB stars in the LMC.  Furthermore, these 13 sources more than double 
the previously known 8 carbon-rich AGB stars in the LMC with $\dot{M}$ higher than 
10$^{-5}$~$M_\odot$~yr$^{-1}$ \citep{vL99,Getal07}.
More intriguingly, these sources typically exceed the maximum $\dot{M}$ expected 
for both oxygen-rich and carbon-rich AGB stars \citep{vL99}.  For example,
the $\dot{M}$ of 054859.98$-$703322.5, 2.3$\times$10$^{-4}$~$M_\odot$~yr$^{-1}$,
is much higher than the maximum expected for its bolometric luminosity,
$4\times10^{-5}$~$M_\odot$~yr$^{-1}$.

Comparisons of carbon stars in the Magellanic Clouds and the Galaxy show that 
$\dot{M}$ does not depend on metallicity \citep{Metal05,Sloan08}, because $\dot{M}$
depends on the amount of carbon available in the atmosphere, which is ultimately 
synthesized within these stars.  These new extreme carbon stars show that the highest 
$\dot{M}$ found in the LMC are comparable to those of their Galactic counterparts 
\citep[e.g.,][]{Speck08}.  The mean bolometric luminosity of the EROs, 
$\sim$7,100~$L_\odot$, corresponds to main-sequence masses of 1.5--2.5\,$M_{\odot}$ 
\citep[using Fig.\,20 in][]{VW93}, where the lower limit is set by the mass range 
to become carbon stars (1.5--5.0\,$M_{\odot}$).  It is puzzling that such low-mass 
stars could reach the highest $\dot{M}$.  This might be caused by some specific  
combinations of abundance and pressure--temperature profile in the stellar 
atmosphere, resulting in an unusual condensation sequence \citep{LF95}.  It is 
also possible that a carbon-rich dust disk is responsible for the SiC absorption
then the $\dot{M}$ based on a spherical symmetry is not valid.  These possibilities
will be considered in greater detail in future work.

\subsection{Spectral Features}

The occurrence of SiC absorption features is surprising.  These features are rare in Galactic 
extreme carbon stars, with only two unambiguous detections \citep{Speck97} 
and another 8 more tentative detections \citep{Petal07,Speck08}.  This raises the 
question of why the SiC absorption feature is seemingly common in the LMC extreme carbon 
stars.

Recent {\it Spitzer} observations show that planetary nebulae (PNe) in the LMC often show 
strong SiC emission \citep{Setal07,Betal08}, while in Galactic PNe, it is generally
weak \citep{B83,Retal02}.
The presence of SiC features in EROs and PNe in the LMC suggests that (a) SiC grains 
are formed more abundantly in the LMC than previous recognized \citep[e.g.,][]{Zijlstra06}; 
and (b) that the LMC metallicity allows SiC to form and survive in the late AGB phase and
into the PN phase more efficiently than in the Galaxy.   It has been hypothesized by
\citet{Leisenring08} that MgS tends to form as a mantle on pre-existing SiC grains and
gives rise to the emission feature at 30~$\mu$m.  This hypothesis is supported by a
recent study of the correlation between mass-loss rate and the amount of SiC required to 
model Galactic extreme carbon star spectra \citep{Speck08}.  Only two of the LMC EROs show 
evidence for a MgS feature at 30~$\mu$m indicating that the SiC grains may not be coated 
for some of these objects.
Current models for the effect of metallicity on condensation sequences in the Galaxy 
and the Magellanic Clouds \citep[e.g.,][]{Lagadec07,Leisenring08,Speck06} need to be 
re-analyzed in light of this discovery.

\acknowledgments 

This research was supported by NASA grants JPL1264494 and JPL1290956.  We thank S.~D.~Points 
for obtaining the ISPI observations used in this paper.  This research has made use of the 
SIMBAD database, operated at CDS, Strasbourg, France.

\end{document}